\documentclass{cupconf}
\usepackage{psfig}

  \checkfont{eurm10}
  \iffontfound
    \IfFileExists{upmath.sty}
      {\typeout{^^JFound AMS Euler Roman fonts on the system,
                   using the 'upmath' package.^^J}%
       \usepackage{upmath}}
      {\typeout{^^JFound AMS Euler Roman fonts on the system, but you
                   dont seem to have the}%
       \typeout{'upmath' package installed. cupconf.cls can take advantage
                 of these fonts,^^Jif you use 'upmath' package.^^J}%
       \providecommand\upi{\pi}%
      }
  \else
    \providecommand\upi{\pi}%
  \fi

  \checkfont{msam10}
  \iffontfound
    \IfFileExists{amssymb.sty}
      {\typeout{^^JFound AMS Symbol fonts on the system, using the
                'amssymb' package.^^J}%
       \usepackage{amssymb}%
       \let\le=\leqslant  \let\leq=\leqslant
       \let\ge=\geqslant  \let\geq=\geqslant
      }{}
  \fi

  \IfFileExists{amsbsy.sty}
    {\typeout{^^JFound the 'amsbsy' package on the system, using it.^^J}%
     \usepackage{amsbsy}}
    {\providecommand\boldsymbol[1]{\mbox{\boldmath $##1$}}}

\newcommand\dynpercm{\nobreak\mbox{$\;$dynes\,cm$^{-1}$}}

\providecommand\bnabla{\boldsymbol{\nabla}}
\providecommand\bcdot{\boldsymbol{\cdot}}
\newcommand\biS{\boldsymbol{S}}
\newcommand\etb{\boldsymbol{\eta}}

\newcommand\Real{\mbox{Re}} 
\newcommand\Imag{\mbox{Im}} 
\newcommand\Rey{\mbox{\textit{Re}}}  
\newcommand\Pran{\mbox{\textit{Pr}}} 
\newcommand\Pen{\mbox{\textit{Pe}}}  
\newcommand\Ai{\mbox{Ai}}            
\newcommand\Bi{\mbox{Bi}}            

\newcommand\ssC{\mathsf{C}}    
\newcommand\sfsP{\mathsfi{P}}  
\newcommand\slsQ{\mathsfbi{Q}} 

\newcommand\hatp{\skew3\hat{p}}      
\newcommand\hatR{\skew3\hat{R}}      

\newsavebox{\astrutbox}
\sbox{\astrutbox}{\rule[-5pt]{0pt}{20pt}}
\newcommand{\astrut}{\usebox{\astrutbox}}

\newcommand\GaPQ{\ensuremath{G_a(P,Q)}}
\newcommand\GsPQ{\ensuremath{G_s(P,Q)}}
\newcommand\p{\ensuremath{\partial}}
\newcommand\tti{\ensuremath{\rightarrow\infty}}
\newcommand\kgd{\ensuremath{k\gamma d}}
\newcommand\shalf{\ensuremath{{\scriptstyle\frac{1}{2}}}}

\newcommand\squart{\ensuremath{{\textstyle\frac{1}{4}}}}
\newcommand\thalf{\ensuremath{{\textstyle\frac{1}{2}}}}
\newcommand\Gat{\ensuremath{\widetilde{G_a}}}
\newcommand\ttz{\ensuremath{\rightarrow 0}}
\newcommand\ndq{\ensuremath{\frac{\mbox{$\partial$}}{\mbox{$\partial$} n_q}}}
\newcommand\sumjm{\ensuremath{\sum_{j=1}^{M}}}
\newcommand\pvi{\ensuremath{\int_0^{\infty}%
  \mskip -33mu-\quad}}

\newcommand\etal{\mbox{\textit{et al.}}}
\newcommand\etc{etc.\ }
\newcommand\eg{e.g.\ }

\title[Planet migration]{Planet migration}

\author[E. W. Thommes and J. J. Lissauer]{Edward W. Thommes$^1$ and
Jack J. Lissauer$^2$} \affiliation{$^1$Astronomy Department,
University of California, Berkeley, CA 94720\\[\affilskip] $^2$Space
Sciences Division, NASA Ames Research Center, Moffett Field, CA 94035}
\pubyear{1996}
\volume{538}
\pagerange{119--126}
\date{?? and in revised form ??}
\begin{document}

\maketitle

\begin{abstract}
A planetary system may undergo significant radial rearrangement during
the early part of its lifetime.  Planet migration can come about
through interaction with the surrounding planetesimal disk and the gas
disk---while the latter is still present---as well as through
planet-planet interactions.  We review the major proposed migration
mechanisms in the context of the planet formation process, in our
Solar System as well as in others.

\end{abstract}

\firstsection 
\section{Introduction}
The word planet is derived from the Greek word ``planetes'', meaning
wandering star.  Geocentric views of the Universe held sway until the
Middle Ages, when Copernicus and Kepler developed a better
phenomenological explanation of planetary wanderings, which with small
modifications has withstood the test of time.  Kepler's first law of
planetary motion states that planets travel along elliptical paths
with one focus at the Sun.  Thus, although planets wander about the
sky, in this model their orbits remain fixed and they do not migrate.
In his physical model of the Solar System, Newton theorized that
planets gradually altered one another's orbits, and he felt compelled
to hypothesize occasional divine intervention to keep planetary
trajectories well-behaved over long periods of time. In the early
1800s, Poisson pointed out that planetary-type perturbations cannot
produce secular changes in orbital elements to second order in the
mass ratio of the planets to the Sun, but Poincare's work towards the
end of the 19th century suggests that the Solar System may be chaotic.
The stability of mature planetary systems is a fascinating topic, but
we shall be concerned herein with the potentially much more rapid
migration of planets during and immediately following the epoch of
their formation.  Modern research into this topic began when Goldreich
and Tremaine (1980) showed that density wave torques could have led to
significant orbital evolution of Jupiter within the protoplanetary
disk on a timescale of a few thousand years, and research accelerated when
giant planets were found much closer to their stars (Mayor and Queloz
1995) than predicted by models of their formation (Lin et al. 1996,
Bodenheimer et al. 2000).

In Section 2, we discuss models for the migration of giant planets
within our own Solar System which may have occurred as the results of
interactions of the planets with one another and with small solid
bodies.  Section 3 summarizes models of the potentially substantial
planetary migration that results from (primarily gravitational)
interactions between a planet and a gaseous protoplanetary disk.  The
predictions of this model are compared to observations of Saturn's
rings and moons in Section 4.  In Section 5, the models are applied to
extrasolar planetary systems.

\section{Migration of Jupiter, Saturn, Uranus, Neptune and Pluto}

In studying the accretion of the giant planets, Fernandez and Ip
(1984) first noted that proto-Uranus and -Neptune could undergo
significant orbital migration due to angular momentum exchange with a
(sufficiently massive) planetesimal disk.  The mechanism operates as
follows: Gravitational stirring by Uranus and Neptune imparts high
eccentricities on the surrounding planetesimals.  Those which acquire
sufficiently small perihelia can be ``handed off'' to the
next-innermost planet, with a resultant gain in angular momentum for
the first planet.  In this way, planetesimals get passed inward from
Neptune to Uranus to Saturn and finally to Jupiter, which is massive
enough to readily eject them from the Solar System.  (The other giant
planets are also massive enough to eject planetesimals from the Solar
System; hovewer, the characteristic timescales for direct ejection are
longer than for passing the planetesimals inwards to the control of
Jupiter.)  It is generally accepted that planetesimal scattering by
the giant planets, principally Jupiter, is what formed the Oort cloud,
the quasi-spherical distribution of comets orbiting at $\sim 10^3-10^5$ AU
from the Sun (Duncan et al. 1987).  Since Jupiter does the lion's
share of the work, it undergoes a net loss of angular momentum and its
orbit shrinks; Saturn, Uranus and Neptune mainly contribute by passing
planetesimals down, so they gain angular momentum.  For a planet on a
circular orbit, change in angular momentum $L$ is related to change in
semimajor axis $a$ according to
\begin{equation}
\Delta L = \frac{1}{2} M \sqrt{\frac{G M_{\odot}}{a}} \Delta a,
\end{equation}
where $M$ is the planet's mass, so Jupiter, being the innermost and most
massive, migrates the shortest distance, while Neptune migrates
farthest. Hahn and Malhotra (1999) investigated this migration
mechanism in detail, using it to try and reproduce the eccentricities
of Pluto and the other Kuiper belt objects sharing the Neptune
exterior 3:2 resonance (Plutinos).  They found that with a planetesimal
disk of about 50 Earth masses (M$_{\oplus}$), Neptune migrates the
right distance to pump the eccentricities of objects carried along in
its 3:2 resonance to the observed values, $e \sim 0.3$.
Larger/smaller disk masses produced too much/not enough migration.
Migration takes place over a timescale of tens of millions of years,
after which time it stalls due to depletion of the planetesimals.



\begin{figure}
\begin{center}
\psfig{figure=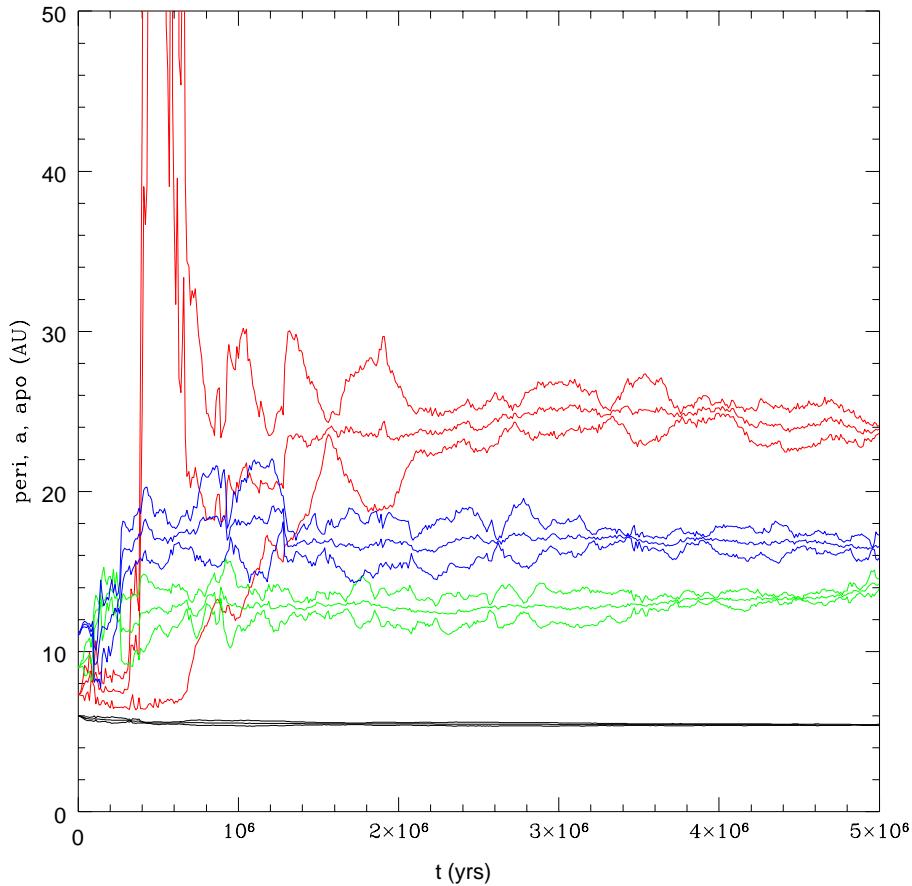,width=5in}
\caption{A numerical simulation showing how Uranus and Neptune could
have originated among Jupiter and Saturn.  Semimajor axis, perihelion
and aphelion distance as a function of time are plotted for each body.
Initially, four 10 M$_{\oplus}$ bodies are placed on circular orbits
in the Jupiter-Saturn region.  The innermost one's mass is increased
over a $10^5$ year timescale to that of Jupiter, to simulate the
accretion of a massive gas envelope.  As a result, the remaining
bodies are scattered outward; their initially high eccentricities are
damped by interaction with the planetesimal disk which exists beyond
$\sim$ 12 AU.  After 5 Myrs of evolution, the result is a giant planet
system which looks qualitatively similar to our own, with the
scattered and recircularized bodies standing in for Saturn, Uranus and
Neptune.  See Thommes et al. (2002a) for details.}
\label{fig1}
\end{center}
\end{figure}

A preceeding phase of migration for Neptune, Uranus and Saturn,
shorter but more violent, was proposed in the model of Thommes et
al. (1999, 2002a).
In trying to account for the formation of the ``ice giants'', Uranus
and Neptune (14.6 and 17.2 M$_{\oplus}$ respectively, $\sim 90\%$ of
which are condensibles) at their present orbital radii (19 and 30 AU
respectively), one runs into serious timescale problems (Lissauer et
al. 1995, Levison and Stewart 2001, Thommes et al. 2002b).  At such
large heliocentric distances, the accretion timescale of such massive
bodies far exceeds the lifetime of the nebular gas (about $10^7$ years
or less, e.g., Strom et al. 1990); once planetesimal random velocities
are no longer damped by aerodynamic gas drag, it may not be possible
to produce an ice giant-sized body on {\it any} timescale.  Indeed, at
20 AU, the mass at which the surface escape velocity from a
(Uranus/Neptune-density) body equals the escape velocity from the Sun,
is only a bit over an Earth mass.  In the model of Thommes et al., the
Jupiter-Saturn region serves as the birthplace of {\it all} the giant
planets, thus alleviating the timescale problem.  Assuming that gas
giant planets form by core accretion (e.g., Pollack et al. 1996), one
of the protoplanets---likely proto-Jupiter---eventually reaches
runaway gas accretion and acquires a massive gas envelope, abruptly
(over $\sim 10^5$ years) expanding its gravitational reach and
destabilizing its neighbours' orbits.  As a result the other giant
protoplanets are scattered, predominantly outwards, ending up with
aphelia in the still accretionally unevolved outer planetesimal disk.
Dynamical friction with the planetesimals then reduces the
eccentricities of these scattered giant protoplanets, decoupling them
from Jupiter and from each other on a timescale of a few million
years.  Numerical simulations show that this sequence of events
commonly results in an outer planetary system similar to our own, with
the scattered protoplanets eventually settling down to nearly circular
orbits of radii comparable to those of Saturn, Uranus and Neptune.  An
example is shown in Fig. \ref{fig1}.  A frequent side effect in the
(stochastic) simulations is strong gravitational stirring of the
Kuiper belt region; this may help to account for the high random
velocities of objects in the ``classical'' Kuiper belt (e.g., Petit et
al. 1999, Kuchner et al. 2002).

{\it In situ} accretion of Uranus and Neptune could still have
occurred if the growing ice giants had somehow continued to be
supplied with planetesimals of relatively low velocity dispersion.
There are several ways in which this could have happened.  One
possibility is that differential migration between the growing
protoplanets and the planetesimals caused the former to sweep,
predator-like, through heretofore dynamically cold parts of the
planetesimal disk (Ward and Hahn 1995, Tanaka and Ida 1999, Bryden et
al. 2000).  Accretion would stall when the protoplanet mass becomes
sufficiently large that it no longer ploughs through the
planetesimals, but instead captures the planetesimals in its
mean-motion resonances and pushes them ahead of it (Kary et al. 1993).
Insofar as the differential migration is to come about from
interaction with the gas disk (aerodynamic gas drag on the
planetesimals, gravitational interaction of the protoplanets with the
disk (see below), or some combination thereof), this mechanism will
only operate for $\sim 10^7$ years.  After that, migration will become
much slower, since migration-inducing interactions would only involve
the much less massive solids component of the protoplanetary disk.
Another possibility is that collisions among the planetesimals acted
to damp their velocities sufficiently for significant further
accretion to take place after dissipation of the gas.  Thommes et
al. (2002b) make a simple estimate of post-gas growth timescales in
the limit of ``maximally effective'' collisional damping (meaning
simply that the random velocity damping timescale is taken to be equal
to the inter-planetesimal collisional timescale), and show that
planetesimals of order ten kilometers in size or smaller---collision
rate is inversely proportional to planetesimal radius---could
conceivably permit the accretion of $\sim$ 10 M$_{\oplus}$ bodies in
the 20 - 30 AU region in of order $10^8 - 10^9$ years.

\section{Disk - Planet Interactions}

A disk with nonzero viscosity will transport angular momentum outward,
resulting in the spreading of the disk; less and less of the disk
material ends up with more and more of the angular momentum and as a
result, there is a net inward transport of mass (Lynden-Bell and
Pringle 1974).  In modeling disk viscosity $\nu$, the prescription of
Shakura and Sunyaev (1973) is commonly used:
\begin{equation}
\nu = \alpha h^2 \Omega,
\label{shak-sun}
\end{equation}
where $\alpha$ is a dimensionless parameter, $h$ is the disk
half-thickness, and $\Omega$ is the disk angular velocity.  Estimates
for $\alpha$ in the protosolar nebula range from $10^{-4}$ to
$10^{-2}$ (e.g., Cabot et al. 1987, Dubrulle 1993).  The source of
viscosity is unknown; it may be turbulent motion resulting from
convective instability in the disk (Lin and Papaloizou 1980), damping
of density waves launched by embedded planetary bodies (Larson 1989,
Goodman and Rafikov 2001), or magnetohydrodynamic turbulence (Balbus
and Hawley 1991).  The latter mechanism requires that the disk be
sufficiently ionized to be strongly coupled to the magnetic field.  At
a radius of less than about 0.1 AU from the star, collisional
ionization ought to be able to accomplish this; further out,
ionization by cosmic rays near the disk surface will dominate.  Thus
it is possible that beyond $\sim$ 0.1 AU, a protoplanetary disk only
transports angular momentum within relatively thin layers on the outer
surfaces of the disk (Gammie 1996).

Individual gas molecules move along paths which are nearly Keplerian
ellipses.  However, the picture becomes more complex once
protoplanets form and perturb the disk.  This problem was first
investigated in the context of satellite interactions with planetary
rings by Goldreich and Tremaine (1980), and extended by Ward (1986,
1997) and Artymowicz (1993) to planetary interactions with a
protoplanetary disk.  A review of the theory of planet-disk
interactions is given by Lin et al. (2000), as well as by Goldreich and
Sari (2002), who investigate the resulting evolution of planet
eccentricities.  We provide an abbreviated summary below.

It is convenient to begin by expressing the gravitational potential of
the planet as a Fourier series:
\begin{equation}
\psi = \Sigma_l \Sigma^{\infty}_{m=0} \psi_{l,m}(r)
\cos[m(\phi-\Omega_{l,m}t)],
\end{equation}
where $\phi$ is the azimuthal angle, $\psi_{l,m}(r)$ is an amplitude
which depends on radius, and $\Omega_{l,m}=\Omega_p+(l-m)\kappa_p/m$
is the pattern speed, with $\Omega_p$ being the angular frequency of a
planet on a circular orbit, and $\kappa_p$ being the planet's
epicyclic (radial oscillation) frequency.  Both $l$ and $m$ are
integers; for each value of the azimuthal mode number $m$, there
are---to lowest order in eccentricity---three components that
contribute to the pattern speed: $\Omega_{l=m}=\Omega_p$, and
$\Omega_{m\pm1,m} = \Omega_p \pm \kappa_p/m$.  Each component, in
turn, has three associated resonances: a corotation resonance, at
which the pattern speed matches the disk angular velocity,
$\Omega_{l,m}=\Omega_d$, and an inner and outer Lindblad resonance, at
which the difference between the disk angular velocity and the pattern
speed is a harmonic of the epicyclic frequency,
$\Omega_d-\Omega_{l,m}=\pm \kappa_d/m$.  For Keplerian orbits,
$\kappa_p=\Omega_p$ and $\kappa_d=\Omega_d$.  At each of the resonance
sites, angular momentum and energy are exchanged between the disk and
the planet, and significant epicyclic motion is excited in the gas.
We can view the situation as a resonating disk particle always being
at the same phase in its radial oscillation when it experiences a
particular phase of a Fourier component of the satellite's forcing.
This enables continued coherent kicks from the satellite to build up
the particle's radial motion, and significant forced oscillations may
thus result.  In this way, spiral density waves---horizontal density
oscillations which result from the bunching of streamlines of gas
molecules on eccentric orbits---are launched.  If the perturbing planet
is inclined relative to the disk, resonances involving the vertical
frequencies will also arise, and these will launch spiral bending
waves, which are vertical corrugations of the disk plane resulting
from the induced coherent inclinations of gas molecule orbits.  We do
not consider vertical resonances further, since they do not directly
affect planet migration; the topic is investigated in detail by, e.g.,
Lubow and Ogilvie (2001).

\section {Saturn's Rings - Observational Tests of Disk-Satellite Interactions}
Several of Saturn's numerous moons orbit near or within the planet's
spectacular main ring system.  These moons excite spiral density waves
at resonant locations within the rings of Saturn.  Gaps (or ring
edges) are produced at strong resonances and close to moons where
resonances overlap.  Pan, a small moon within the A
ring, has cleared a gap around its orbit, and Pandora and Prometheus,
somewhat larger moons orbiting just exterior Saturn's main ring
system, confine particles to the narrow, irregular F ring. Spiral
density waves and sharp ring boundaries were detected in Saturn's
rings by four instruments on the Voyager spacecraft.  These data
provide for valuable tests of models of interactions between
secondaries and astrophysical disks.
 
\subsection{Wave characteristics and ring properties}
Several dozen density waves in Saturn's rings have thus far been
identified with exciting resonances and analyzed to determine the
local surface mass density of the rings (e.g., Esposito et al. 1984,
Rosen et al. 1991).  The observed and predicted resonance locations
agree within observational uncertainties, which are generally less
than 1 part in $10^4$.  The surface density, $\sigma$, at most wave
locations in the optically thick A and B rings is of order 50
g/cm$^2$.  Measured values in the optically thin C ring are $\sigma
\sim$ 1 g/cm$^2$; an intermediate value of $\sigma \sim$ 10 g/cm$^2$
has been estimated for Cassini's Division.

The outer edges of the B and A rings are maintained by the Mimas 2:1
and Janus 7:6 resonances, which are the strongest resonances within
the ring system (Smith et al. 1981, Holberg et al. 1982, Porco et
al. 1984a, Lissauer and Cuzzi 1982, Borderies et al.
1982).  Nearly empty gaps with embedded optically thick ringlets have
been observed at strong resonances located in optically thin regions
of the rings (Holberg et al. 1982).  These features are probably
caused by a resonance-related process; however, no explanation for the
embedded ringlets currently exists.  Nearly empty gaps with embedded
ringlets have also been observed at nonresonant locations (Porco et
al. 1984b).

Although resonance features in Saturn's rings are well understood in
many respects, there remain several major outstanding issues.  Some of
these problems are related to angular momentum transport, a key factor
in planetary migration.

     The damping behavior of most observed spiral waves differs
significantly from that predicted by the simple fluid approximation in
a constant viscosity disk (Goldreich and Tremaine 1978).  This
has led to severe problems with attempts to estimate the viscosity of
the rings using observations of wave damping.  Some of the factors
which can influence the damping behavior of waves are variations in
the background surface density and velocity dispersion, interference
with other waves and wave nonlinearities (Lissauer et al. 1984).  The
damping of nonlinear density waves has been studied in detail by Shu
et al. (1985), who find that damping rates can be very sensitive to
particle collision properties and to optical depth.  Thus, the
anomalous damping behavior of many spiral waves presents both a
challenge and an opportunity for researchers attempting to deduce ring
properties other than surface mass density from the study of nonlinear
waves (Longaretti and Borderies 1986).  Spiral waves carry (positive
or negative) angular momentum, and deposit it in regions of the disk
in which they damp, so the poor correspondence between theory and
observation is of concern to planet migration modelers.

     The cause of the observed enhancement of material in regions
where strong waves propagate is poorly understood.  Density waves
excited at inner Lindblad resonances carry negative angular momentum,
i.e., the angular momentum of the ring particles is temporarily
reduced by the passage of these waves.  When such waves damp,
particles drift inwards.  Resonant removal of angular momentum causes
sharp outer ring edges and gaps to be produced at the strongest
resonances within the ring system (Borderies et al. 1982).  Waves are
observed to be excited at the strongest resonances which do not
produce gaps.  However, waves do not appear to deplete material from
the regions in which they propagate; on the contrary, a surface
density enhancement is often observed in such regions (Holberg et
al. 1982, Longaretti and Borderies 1986, Rosen et al. 1991).  

The problem of ``dredging'' of ring material due to wave damping has
never been solved in a self-consistent manner, in which the evolution
of a region due to wave propagation, wave damping and general ring
viscosity is following until a quasi-steady state is attained.  (The
most detailed study of this problem thus far attempted is presented by
Borderies et al. 1986.)  Damping of the wave in the outer portion of
the region in which it propagates could bring material towards
resonance, but what stops (or at least slows) this material from
further inward drift?  The answer to this question may be relevant to
two of the major questions facing ring theorists today: What maintains
inner edges of the major rings of Saturn? and Do the short timescales
for ring evolution due to density wave torques imply Saturn's rings
are much younger than the planet itself?

\subsection{Migration of moons}
Before Voyager arrived at Saturn, Goldreich and Tremaine (1978)
predicted that torques due to density waves excited by the moon Mimas
at its 2:1 resonance had removed sufficient angular momentum from ring
material to have cleared out Cassini's Division, a 4000 km wide region
of depressed surface mass density located between the broad high
density A and B rings.  Although the hypothesis of density waves
clearing Cassini's Division remains unverified, Voyager found a
multitude of density waves excited by small newly-discovered
satellites orbiting near the rings.  Although the torque at these
individual resonances is less than that at Mimas' 2:1 resonance, the
sum of their torques is much greater.  Moreover, the waves are
observed and amplitudes agree with theory to within a factor of order
unity.

     Goldreich and Tremaine (1982) pointed out that the back torque
the rings exert on the inner moons causes them to recede on a
timescale short compared to the age of the Solar System; estimates
suggest that all of the small moons orbiting inside the orbit of Mimas
should have been at the outer edge of the A ring within the past
10$^8$ years, with Prometheus' journey outward occurring on a
timescale of only a few millions years; more recent, lower estimates
for the masses of these moons increase these timescales by a factor of
a few, but do not solve the problem.  Resonance locking to outer more
massive moons could slow the outward recession of the small inner
moons; however, angular momentum removed from the ring particles
should force the entire A ring into the B ring in $\sim$ 10$^9$ yr
(Lissauer et al. (1984) and Borderies et al. (1984) quote somewhat
shorter times based on larger masses of the moons).

If the calculations of torques are correct, and if no currently
unknown force counterbalances them, then small inner moons and/or the
rings must be new, i.e., much younger than the age of the Solar
System.  Both of these possibilities appear to be {\it a priori}
highly unlikely.  Rings could be remnants of Saturn's protosatellite
disk which never accreted into moon-sized bodies due to the strong
tidal forces of Saturn inside Roche's limit, in which case they would
be $\sim 4.5 \times 10^9$ yr old.  Alternatively, Saturn's ring
particles could be part of the debris from a moon that was
collisionally disrupted, in which case they would most likely date
from the first $\sim 10^9$ years of the Solar System, when much more
debris large enough to cause such a disruption was available than is
today (Lissauer et al. 1988), or from a tidally disrupted giant comet,
although this possibility is also {\it a priori} unlikely (Dones
1991).  Recent accretion of the inner moons within the ring system may
be possible (Borderies et al. 1984), but why did such accretion only
occur during the past $\sim$ 10$^8$ yr?  For these reasons, the issue
of short timescales due to density wave torques is a major outstanding
problem in the field of planetary rings.


\section{Migration of Extrasolar Planets}

\subsection{Type I migration}
Goldreich and Tremaine (1980) pointed out that interactions with the
gas disk could move planets large distances during the lifetime of the
gas.
Relative to the location of the planet, the inward/outward-propagating
density waves carry a net negative/positive flux of angular momentum.
If this angular momentum is deposited in the disk (as opposed to
reflecting at the disk boundaries and returning to the planet), there
will be a positive/negative torque on the planet from the
interior/exterior parts of the disk.  Variations in disk properties
ought, in general, to produce a mismatch between the net interior and
exterior torques, and a resultant migration rate for the planet of
order
\begin{equation}
v_{\rm I}=k_1 \frac{M}{M_*}r \Omega \frac{\Sigma_d r^2}{M_*}\left (\frac{r
\Omega}{c} \right )^3,
\label{Type I migration rate}
\end{equation}
where $k_1$ is a measure of the torque asymmetry, $M$ is the mass of
the planet, $M_*$ is the mass of the primary, $r$ is the distance from
the primary, $\Sigma_d$ is the disk surface density, and $c$ is the
gas sound speed.  From dimensional arguments, $k_1$ should scale with
the disk aspect ratio $h/r$ (Goldreich and Tremaine 1980).  This type
of orbital drift is commonly referred to as Type I migration.  Ward
(1997) showed that the outer torque ought to dominate in general,
resulting in orbital decay for the planet.

\subsection{Gap opening and Type II migration}

The migration rate continues to increase linearly with planet mass as per
Eq. (\ref{Type I migration rate}), until the torque saturates and the
planet pushes the inner and outer parts of the disk apart, opening a
gap for itself.  Having thus established a (perhaps imperfect) barrier
to the passage of disk material across its orbit, it is then (more or
less) locked to the viscous evolution of the disk, i.e., carried
inward along with the net inward flow of the disk material---assuming,
of course, that this is all happening in a part of the disk with
nonzero viscosity.  This type of orbital evolution is called Type II
migration.  The characteristic velocity for this mode of migration is
\begin{equation}
v_{\rm II} = k_2 \alpha  r \Omega\left (\frac{c}{r \Omega} \right )^2,
\label{Type II}
\end{equation}
where $k_2$ is a constant of order unity.

It is unclear when the formation of a gap occurs; this depends on both
the viscosity of the disk, and on the way in which the density waves
are dissipated.  Lin and Papaloizou (1993) required the density waves
to be strongly nonlinear as soon as they are launched, so that they
immediately shock and deposit their angular momentum in the disk.
They showed that the minimum planetary mass to accomplish this, in the
limit of zero disk viscosity, is such that the Hill (or Roche) radius
of the planet, defined as
\begin{equation}
r_{\rm H} = \left ( \frac{M}{3 M_*} \right )^{1/3} r,
\label{rhill}
\end{equation}
is equal to the half-thickness of the disk, $h$.  This
criterion---often called the thermal condition---yields a gap-opening
mass of 
\begin{equation}
M_{\rm thermal}^{\rm crit} \sim M_* \left ( \frac{h}{r} \right )^3 \sim
100 \left ( \frac{r}{\rm 1 AU} \right )^{3/4} {\rm M_{\oplus}},
\label{thermal}
\end{equation}
where the numerical estimate is obtained using a Solar-mass primary
and the standard Hayashi (1981) nebula model, which has a
half-thickness of
\begin{equation}
h=0.0472(a/{\rm 1 AU})^{-5/4}{\rm AU},
\end{equation}
and a minimum surface density---obtained from smoothly spreading out
the mass contained in the planets and enhancing the gas content to
solar abundance, of
\begin{equation}
\Sigma_{\rm min} = 1.7 \times 10^3 \left ( \frac{a}{\rm 1 AU}\right )^{-3/2}{\rm g/cm}^2.
\end{equation}
In a disk with nonzero viscosity, an additional condition for
maintaining a gap is that the rate of angular momentum transfer
across the gap exceed the intrinsic viscous angular momentum transport
rate of the disk (Lin and Papaloizou 1993, Bryden et al. 1999).  This
leads to a critical mass of{\rm 
\begin{equation}
M^{\rm crit}_{\rm viscous}\sim 40 M_* \alpha \left ( \frac{h}{r}\right
)^2 \sim 300 \left ( \frac{\alpha}{10^{-2}}\right ) \left
( \frac{r}{\rm 1 AU}\right )^{1/2}{\rm M_{\oplus}}.
\label{viscous}
\end{equation}
In a viscous disk, the minimum gap opening mass is then Min$[M_{\rm
thermal}^{\rm crit}, M^{\rm crit}_{\rm viscous}]$.  

On the other hand, Ward and Hourigan (1989) assumed that damping is
linear and independent of the perturber's mass.  In this way, they
obtained a much smaller critical mass of 
\begin{equation}
M^{\rm crit}_{\rm inertial} \sim \frac{h^3 \Sigma}{r} \sim 0.007 \left
( \frac{r}{\rm 1 AU}\right )^{5/4} {\rm M_{\oplus}},
\label{stallmass_ward}
\end{equation}
not for gap-opening, but for the planet to induce a density contrast
in the disk---essentially to pile up disk mass ahead of itself---that
is strong enough to stall its migration.  They referred to this as the
inertial mass.  

The result of Rafikov (2002) is intermediate between $M^{\rm
crit}_{\rm thermal}$ and $M^{\rm crit}_{\rm intertial}$.  In this
model, density waves travel some distance in the disk before
dissipating via weak nonlinearity.  The critical mass obtained in this
way is around 2 M$_{\oplus}$ at 1 AU in an inviscid minimum-mass disk.

\subsection{Migration versus formation?}
Type I migration timescales can becomes as short as $\sim 10^4$ years
for a planet of mass $\sim 10\, {\rm M_{\oplus}}$ in a sufficiently
viscous disk (Ward 1997), two to three orders of magnitude less than
the lifetime of the gas disk.  In the core accretion model, the
critical mass needed to initiate runaway gas accretion is thought to
be of order 10 ${\rm M_{\oplus}}$; the formation timescale for a body
of this size is of order a million years or more (e.g., Lissauer 1987,
Lissauer and Stewart 1993, Weidenschilling 1998, Kokubo and Ida 2000,
Thommes et al. 2002b).  So, unless gap formation/stalling occurs at
masses far below 10 M$_{\oplus}$, a growing core would spiral into its
parent star long before it got large enough to accrete a massive
atmosphere.  It has been proposed that rapid migration could actually
speed up the accretion process (Ward 1986), in the same way described
above for the formation of Uranus and Neptune.  However, since
formation timescales are shorter at smaller radii, an inward-migrating
protoplanet is likely to encounter not a pristine planetesimal disk,
but instead the stirred-up remains of a disk which has itself already
produced large protoplanets.  Furthermore, simulations have shown that
even in the idealized case, accretion efficiency is low and the disk
has to be enhanced by at least a factor of five relative to minimum
mass in order to allow a protoplanet starting at 10 AU to grow to
$\sim 10\, {\rm M_{\oplus}}$ before it falls into the star (Tanaka and
Ida 1999).

Clearly, rapid Type I migration constitutes a major potential problem
for the core accretion model of giant planet formation.  However,
there may be other ways out, apart from the possibility that this
phase ends quickly.  For one thing, the analysis thus far has been
restricted to a single body interacting with the disk; it is unclear
whether the coupling to the disk remains as strong when multiple
protoplanets are launching density waves in close proximity to each
other.  Also, Tanaka et al. (2002) demonstrated that
reflection of the density waves at the outer edge of the disk could
permit a non-migrating steady state to be attained without the
formation of a gap.  

\subsection{Migration and the properties of extrasolar planets}
Even after locking themselves in a gap, giant planets are likely to
undergo significant migration (e.g., Ward 1997).  
The ensemble of extrasolar planets detected by radial velocity
searches (Mayor and Queloz 1995, Marcy et al. 2000) provides
considerable observational support for migration playing a large role
in the formation of planetary systems.  The most direct clue is the
large number planets on close-in orbits.  Nearly half of the known
planets orbit closer than 1 AU to their parent star.  Although there
is certainly a strong selection effect favouring detection of
short-period planets, it is clear that a non-negligible fraction of
giant planets somehow end up on such orbits.  {\it In situ} formation
is one possibility, but there are a number of problems in trying to
construct such models; in particular, a very high surface density of
solids and/or substantial planetesimal migration would be required to
form a giant planet core-sized body (Bodenheimer et al. 2000).
Formation at larger stellocentric distances followed by inward
migration seems to provide a more natural explanation for planets like
those orbiting 51 Peg and $\rho$ CrB.

Migration may also have facilitated interactions between planets in
some of the detected multiple-planet systems.  The two planetary
companions of Gliese 876 are in a 2:1 mean-motion resonance with each
other (Marcy et al. 2001).  Lee and Peale (2002) showed that capture
into this resonance would occur if the two planets were originally
farther apart in orbital period, and were induced to migrate toward
each other, assuming their eccentricities remained low throughout ($e
\sim 10^{-2}$ or less).  Bryden et al. (2000) and Kley (2000) performed
hydrodynamic simulations of two gap-opening planets embedded in a gas
disk, and demonstrated that, if the planets' orbits are sufficiently
close together, they will tend to clear the annulus of gas between
them on timescales as short as thousands of years.  The result is that
both planets end up sharing a gap, and are pushed toward each other by
the inner and outer parts of the disk.  Aside from the Gliese system,
the two planets discovered in HD 82943 
appear to also be in a 2:1 mean motion resonance; numerical
simulations suggest that, given the inferred orbital parameters, only
resonant configurations are stable (Jianghui et al. 2002).
Furthermore, the periods of the inner two companions of the putative
three-planet system 55 Cnc are very close to a 3:1 commensurability
(Marcy et al. 2002).  Capture into this resonance, and other
higher-order resonances, can occur if initial eccentricities of the
convergently migrating planets are nonzero.

{\it Divergent} migration in multiple-planet systems is another
possible mechanism for reproducing some of the observed properties of
extrasolar planets.  Chiang et al. (2002) show that if
planets on initially circular orbits move apart in orbital period, the
noncapturing resonance passages they encounter can induce significant
eccentricities in the orbits of both bodies.  Such a mechanism may
help to account for the generally large eccentricities of those
extrasolar planets with orbital radii greater than a few hundredths of
an AU, out of range for tidal circularization by their parent star
(e.g., Lin et al. 2000).  Divergent migration may come about if there is
a sufficiently steep gradient in disk viscosity, so that the inner
planet migrates faster than the outer one.  In particular, with
reference to the model of Gammie (1996), if one planet is interior to
the collisional ionization radius, where viscosity will be high, and
the other is exterior to it, where viscosity is low, divergent
migration may result.  However, it is required that the gas annulus
between the two planets persist, notwithstanding the results cited
above.

\subsection{Type III migration}
In a low-mass disk---one in which the giant planet mass is comparable
to or greater than the disk mass---the planet's inertia will slow the
evolution of the disk.  Thus the planet's migration speed will be
inversely proportional to its mass:
\begin{equation}
v_{\rm III} \sim \frac{M_{disk}}{M_{planet}+M_{disk}} v_{\rm II}
\end{equation} 
At the same time, accretion of
disk material onto the star will trail off as the part of the disk
interior to the planet drains onto the star.  This mode can be
referred to as Type III migration.  Observations hint that something
like this may be happening in TW Hydra, a ten million year old T Tauri
star (Calvet et al. 2002).  TW Hydra's accretion rate is very low
compared to younger T Tauri stars, and at the same time, the spectral
energy distribution of its infrared excess is best fit by a disk with
a sharp inner edge at 4 AU, perhaps signifying truncation by a giant
planet.  Interestingly, however, TW Hydra's disk mass is estimated to
be quite large, $\sim 0.6 M_*$.  A 10 million year old disk still
containing this much mass implies a low disk viscosity, whether the
disk is simply accreting slowly, or whether it is indeed being kept at
bay by a planet.  Calvet et al.
estimate $\alpha < 10^{-3}$.


\section{Conclusions} 

The longstanding view of planet formation as an orderly process,
involving little radial migration of material, is being made to look
increasingly inaccurate by both observational and theoretical
findings.  In our own Solar System, the high eccentricities of objects
in exterior mean-motion resonances with Neptune imply an outward
migration of several AU by the outer ice giant; modeling suggests that
Uranus, and to a lesser degree Jupiter and Saturn, likewise underwent
migration as they cleared the surrounding planetesimal disk.  An
earlier and more violent period of migration could have occured if
Uranus and Neptune originally formed among proto-Jupiter and -Saturn;
such a model would alleviate the longstanding formation timescale
problem of Uranus and Neptune, and could simultaneously help to
account for the gravitationally stirred-up state of the Kuiper belt.

In the first ten million years or so of a planetary system's life, the
nebular gas is still present and provides a much larger sink/source of
angular momentum than the planetesimal disk.  Growing protoplanets
exchange angular momentum by launching density waves at resonance
sites in the disk.  From theoretical consideration, this ought to
bring about a net loss of angular momentum and rapid orbital
decay---Type I migration---for bodies of order a few Earth masses.
For a sufficiently massive body, the torques between it and the disk
will be strong enough to open a gap, thus locking the body into the
subsequent viscous evolution of the disk in what is called the Type II
mode of migration.  The core accretion model of giant planet formation
seems to require that the Type I to II transition occur at small
masses, otherwise growing cores will spiral into the star before they
can acquire a massive envelope.  Alternatively, it is possible that
planet formation simply is an enormously wasteful process, which dumps
a steady stream of growing protoplanets onto the primary, and the end
result is whatever happens to be left over when the gas fades away .
This is sometimes called the ``last of the Mohicans'' scenario.
Significant post-formation migration is quite likely responsible for
the large number of planets detected on close-in orbits (``giant
Vulcans'', also referred to as ``hot Jupiters'').  In multiple-planet
systems, convergent and divergent migration of planets can be invoked
to explain, respectively, resonant capture and eccentricity
excitation.  As the nebular gas dissipates, it is likely that the
tables are eventually turned; the planets, heretofore at the mercy of
the gas, assert themselves and serve as anchors to slow down the
viscous evolution of the last remains of the disk, so that migration
ends in a Type III phase.  Clearly, the present observational and
theoretical ``state of the art'' still requires us to use a liberal
amount of conjecture in attempting to sketch a coherent picture of
planet migration.  However, it seems equally clear that migration is
intimately linked with the formation of planetary system, and a
complete picture of the latter will require a full understanding of
the former.



\begin{thebibliography}{}


\bibitem[Artymowicz(1993)]{1993ApJ...419..155A} 
\textsc{Artymowicz, P.} 1993 
On the Wave Excitation and a Generalized Torque Formula for Lindblad 
Resonances Excited by External Potential. \emph{ApJ} \textbf{419}, 155. 

\bibitem[Bodenheimer et al.(2000)]{2000Icar..143....2B} 
\textsc{Bodenheimer, P., 
Hubickyj, O. \& Lissauer, J. J.} 2000
Models of the in Situ Formation of Detected Extrasolar Giant Planets.
\emph{Icarus} \textbf{143}, 2-14. 

\bibitem[Borderies et al.(1982)]{1982Natur.299..209B} 
\textsc{Borderies, N., Goldreich, P. \& Tremaine S.} 1982
 Sharp edges of planetary rings. 
\emph{Nature} {\bf 299}, 209-211. 

\bibitem[Borderies et al.(1984)]{1984prin.conf..713B} 
\textsc{Borderies, N., Goldreich, P. \& Tremaine, S.} 1984
Unsolved problems in planetary ring dynamics.
\emph{IAU Colloq. {\bf 75}: Planetary Rings}, 713-734. 

\bibitem[Borderies et al.(1986)]{Bord86} \textsc{Borderies, N., Goldreich, P. \& Tremaine, S.} 1986 Nonlinear density waves in planetary rings. \emph{Icarus} {\bf 68}, 522-533.

\bibitem[Bryden et al.(1999)]{1999ApJ...514..344B} 
\textsc{Bryden, G., Chen X., 
Lin, D. N. C., Nelson, R. P. \& Papaloizou, J. C. B.} 1999
Tidally Induced Gap Formation in Protostellar Disks: Gap Clearing and Suppression of 
Protoplanetary Growth.
\emph{ApJ} {\bf 514}, 344-367. 

\bibitem[Bryden et al.(2000)]{2000ApJ...544..481B} 
\textsc{Bryden, G., Lin, D. N. C. \& Ida, S.} 2000
Protoplanetary Formation. I. Neptune. \emph{ApJ}
{\bf 544}, 481-495. 

\bibitem[Bryden et al.(2000)]{2000ApJ...540.1091B} 
\textsc{Bryden, G., R{\' o}{\. z}yczka, M., Lin, D. N. C. \& Bodenheimer, P.} 2000
On the Interaction between Protoplanets and Protostellar Disks.
\emph{ApJ} {\bf 540}, 1091-1101. 

\bibitem[Calvet et al.(2002)]{2002ApJ...568.1008C} 
\textsc{Calvet, N., D'Alessio, P., Hartmann, L., Wilner, D., Walsh, A. \& Sitko, M.} 2002 
Evidence for a Developing Gap in a 10 Myr Old Protoplanetary Disk.
\emph{ApJ} {\bf 568}, 1008-1016. 

\bibitem[Cabot et al.(1987)]{1987Icar...69..387C} 
\textsc{Cabot, W., Canuto, V. M., 
Hubickyj, O. \& Pollack, J. B.} 1987
The role of turbulent convection in the primitive solar nebula. I - Theory. II - Results.
\emph{Icarus} {\bf 69}, 387-457. 

\bibitem[Chiang et al.(2002)]{2002ApJ...564L.105C} 
\textsc{Chiang, E.~I., Fischer, D. \& Thommes, E. W.} 2002
 Excitation of Orbital Eccentricities of 
Extrasolar Planets by Repeated Resonance Crossings. \emph{ApJ} {\bf 564}, L105-L109. 

\bibitem[Dones(1991)]{1991Icar...92..194D} 
\textsc{Dones, L.} 1991
A recent cometary origin for Saturn's rings?
\emph{Icarus} {\bf 92}, 194-203. 

\bibitem[Dubrulle(1993)]{1993Icar..106...59D} 
\textsc{Dubrulle, B.} 1993 
Differential rotation as a source of angular momentum transfer in the solar 
nebula.
\emph{Icarus} {\bf 106}, 59. 

\bibitem[Duncan et al.(1987)]{1987AJ.....94.1330D} 
\textsc{Duncan, M., Quinn, T. \& Tremaine S.} 1987
 The formation and extent of the solar system comet 
cloud.
\emph{AJ} {\bf 94}, 1330-1338. 

\bibitem[Esposito et al.(1984)]{1984satn.book..463E} 
\textsc{Esposito, L.~W., Cuzzi, J. N., Holberg, J. B., Marouf, E. A., Tyler, G. L. \& Porco, C. C.} 1984
Saturn's rings - Structure, dynamics, and particle properties. 
\emph{Saturn} 463-545. 

\bibitem[Fernandez and Ip(1984)]{1984Icar...58..109F} 
\textsc{Fernandez, J.~A. \& Ip, W.-H.} 1984
Some dynamical aspects of the accretion of Uranus and 
Neptune - The exchange of orbital angular momentum with planetesimals.
\emph{Icarus} {\bf 58}, 109-120. 

\bibitem[Gammie(1996)]{1996ApJ...457..355G} 
\textsc{Gammie, C.~F.} 1996
 Layered Accretion in T Tauri Disks.
 \emph{ApJ} {\bf 457}, 355. 

\bibitem[Goldreich and Tremaine(1978)]{1978Icar...34..227G} 
\textsc{Goldreich, P. \& Tremaine, S. D.} 1978
The velocity dispersion in Saturn's rings. 
\emph{Icarus} {\bf 34}, 227-239. 

\bibitem[Goldreich and Tremaine(1980)]{1980ApJ...241..425G} 
\textsc{Goldreich, P. \& Tremaine, S.} 1980
 Disk-satellite interactions.
\emph{ApJ} {\bf 241}, 425-441. 

\bibitem[Goldreich and Tremaine(1982)]{1982ARA&A..20..249G}
\textsc{Goldreich, P. \& Tremaine, S.} 1982
 The dynamics of planetary rings.
\emph{ARA \& A} {\bf 20}, 
249-283. 

\bibitem[Goldreich and Sari(2002)]{gs02} 
\textsc{Goldreich, P. \& Sari, R.} 2002
Eccentricity evolution for planets in gaseous disks.
Submitted to \emph{ApJ}.

\bibitem[Hahn and Malhotra(1999)]{1999AJ....117.3041H} 
\textsc{Hahn, J.~M. \& Malhotra, R.} 1999
 Orbital Evolution of Planets Embedded in a Planetesimal 
Disk.
\emph{AJ} {\bf 117}, 3041-3053. 

\bibitem[Hayashi(1981)]{1981PThPS..70...35H} 
\textsc{Hayashi, C.} 1981
Structure 
of the solar nebula, growth and decay of magnetic fields and effects of 
magnetic and turbulent viscosities on the nebula.
\emph{Progress of Theoretical 
Physics Supplement} {\bf 70}, 35-53. 

\bibitem[Holberg et al.(1982)]{1982Natur.297..115H} 
\textsc{Holberg, J.~B., Forrester, W. T. \& Lissauer, J. J.} 1982
Identification of resonance 
features within the rings of Saturn.
\emph{Nature} {\bf 297}, 115-120. 

\bibitem[]{jian} 
\textsc{Jianghui, J. et al.} 2002
 The dynamics of the HD 82943
planetary system. Submitted to \emph{ApJ}.

\bibitem[Kary et al.(1993)]{1993Icar..106..288K} 
\textsc{Kary, D.~M., Lissauer, J. J. \& Greenzweig, Y.} 1993
Nebular gas drag and planetary accretion.
\emph{Icarus} {\bf 106}, 288-307. 

\bibitem[Kley(2000)]{2000MNRAS.313L..47K} 
\textsc{Kley, W.} 2000
 On the migration of a system of protoplanets.
\emph{MNRAS} {\bf 313}, L47-L51. 

\bibitem[Kuchner et al.(2002)]{2002AJ....124.1221K} 
\textsc{Kuchner, M.~J., Brown, M. E. \& Holman, M.} 2002 
Long-Term Dynamics and the Orbital 
Inclinations of the Classical Kuiper Belt Objects.
\emph{AJ} {\bf 124}, 1221-1230. 

\bibitem[Larson(1989)]{1989feps.meet...31L} 
\textsc{Larson, R.~B.} 1989
 The evolution of protostellar disks. In \emph{The Formation and Evolution of Planetary 
Systems} 31-48. 

\bibitem[Lee and Peale(2002)]{2002ApJ...567..596L} \textsc{Lee, M. H. \& 
Peale, S. J.} 2002
Dynamics and Origin of the 2:1 Orbital Resonances of the 
GJ 876 Planets.
\emph{ApJ} {\bf 567}, 596-609. 

\bibitem[Levison and Stewart(2001)]{2001Icar..153..224L} \textsc{Levison, H.~F. \& 
Stewart, G. R.} 2001.
Remarks on Modeling the Formation of Uranus and 
Neptune.
\emph{Icarus} {\bf 153}, 224-228. 

\bibitem[Lin and Papaloizou (1980)]{LP80} \textsc{Lin, D.~N.~C.~ \&
Papaloizou, J. C. B.} 1980. On the structure and evolution of the
primordial solar nebula. \emph{MNRAS} {\bf 191}, 37-48.

\bibitem[Lin and Papaloizou(1993)]{1993prpl.conf..749L} \textsc{Lin, D.~N.~C. \&
Papaloizou, J. C. B.} 1993
On the tidal interaction between protostellar 
disks and companions.
In \emph{Protostars and Planets III}, 749-835. 

\bibitem[Lin et al.(1996)]{1996Natur.380..606L} \textsc{Lin, D.~N.~C., 
Bodenheimer, P. \& Richardson, D. C.} 1996
 Orbital migration of the 
planetary companion of 51 Pegasi to its present location.
\emph{Nature} {\bf 380}, 
606-607. 

\bibitem[Lin et al.(2000)]{2000prpl.conf.1111L} \textsc{Lin, D.~N.~C., 
Papaloizou, J. C. B., Terquem, C., Bryden, G. \& Ida, S.} 2000
 Orbital 
Evolution and Planet-Star Tidal Interaction.
In \emph{Protostars and Planets IV}, 
1111. 

\bibitem[Lissauer and Cuzzi(1982)]{1982AJ.....87.1051L} \textsc{Lissauer, J.~J. \&
Cuzzi, J. N.} 1982
 Resonances in Saturn's rings.
\emph{AJ} {\bf 87}, 1051-1058. 

\bibitem[Lissauer et al.(1984)]{1984plri.coll..385L} \textsc{Lissauer, J.~J., Shu, F. H. \& Cuzzi, J. N.} 1984
 Viscosity in Saturn's rings. In \emph{CNES  
Planetary Rings}, 385-392. 
     
\bibitem[Lissauer et al.(1987)]{L87} \textsc{Lissauer,
J. J.} 1987. Timescales for planetary accretion and the structure of
the protoplanetary disk. \emph{Icarus} {\bf 69}, 249-265.

\bibitem[Lissauer et al.(1988)]{1988LPI....19..683L} \textsc{Lissauer, J.~J., 
Squyres, S. W. \& Hartmann, W. K.} 1988
 Bombardment History of the Saturn 
System. \emph{J. Geophys. Res.} {\bf 93}, 13776-13804. 

\bibitem[Lissauer and Stewart(1993)]{1993prpl.conf.1061L} \textsc{Lissauer, 
J.~J. \& Stewart, G. R.} 1993
 Growth of planets from planetesimals.
In \emph{Protostars and Planets III}, 1061-1088. 

\bibitem[liss95]{} \textsc{Lissauer, J.J., Pollack, J. B., Wetherill, G. W. \& Stevenson, D. J.} 1995
Formation of  the Neptune System.  In \emph{Neptune and Triton},
37-108.

\bibitem[Longaretti and Borderies(1986)]{1986Icar...67..211L} \textsc{Longaretti, 
P.-Y. \& Borderies, N.} 1986
 Nonlinear study of the Mimas 5:3 density wave.
\emph{Icarus} {\bf 67}, 211-223. 

\bibitem[Lubow and Ogilvie(2001)]{2001ApJ...560..997L} \textsc{Lubow, S.~H. \& 
Ogilvie, G. I.} 2001
Secular interactions between inclined planets and a gaseous Disk.
\emph{ApJ} {\bf 560}, 997-1009. 

\bibitem[Lynden-Bell and Pringle(1974)]{1974MNRAS.168..603L} \textsc{Lynden-Bell, 
D. \& Pringle, J. E.} 1974
 The evolution of viscous discs and the origin of the nebular variables.
\emph{MNRAS} {\bf 168}, 603-637. 

\bibitem[Marcy et al.(2000)]{2000prpl.conf.1285M} \textsc{Marcy, G.~W., 
Cochran, W. D. \& Mayor, M} 2000
Extrasolar Planets around Main-Sequence Stars.
In \emph{Protostars and Planets IV},
1285. 

\bibitem[Marcy et al.(2001)]{2001ApJ...556..296M} \textsc{Marcy, G.~W., 
Butler, R. P., Fischer, D., Vogt, S. S., Lissauer, J. J. \& Rivera, E. J.} 
2001
 A Pair of Resonant Planets Orbiting GJ 876.
\emph{ApJ} {\bf 556}, 296-301. 

\bibitem[]{marcy02} \textsc{Marcy et al.} 2002
A Planet at 5 AU Around 55 Cancri. Submitted to \emph{ApJ}. 

\bibitem[Mayor and Queloz(1995)]{1995Natur.378..355M} \textsc{Mayor, M. \& 
Queloz, D.} 1995
 A Jupiter-Mass Companion to a Solar-Type Star.
\emph{Nature} {\bf 378}, 355. 

\bibitem[Petit et al.(1999)]{1999Icar..141..367P} \textsc{Petit, J., Morbidelli, A. 
\& Valsecchi, G. B.} 1999 
Large Scattered Planetesimals and the Excitation of the Small Body Belts.
\emph{Icarus} {\bf 141}, 367-387. 

\bibitem[Pollack et al.(1996)]{1996Icar..124...62P} \textsc{Pollack, J.~B., 
Hubickyj, O., Bodenheimer, P., Lissauer, J. J., Podolak, M. \& Greenzweig, Y.} 
1996
 Formation of the Giant Planets by Concurrent Accretion of Solids and Gas.
\emph{Icarus} {\bf 124}, 62-85. 

\bibitem[Porco et al.(1984a)]{1984Icar...60...17P} \textsc{Porco, C., 
Danielson, G. E., Goldreich, P., Holberg, J. B. \& Lane, A. L.} 1984a.
 Saturn's nonaxisymmetric ring edges at 1.95 $R(s)$ and 2.27 $R(s)$.
\emph{Icarus} {\bf 60}, 17-28. 

\bibitem[Porco et al.(1984)]{1984Icar...60....1P} \textsc{Porco, C., 
Nicholson, P. D., Borderies, N., Danielson, G. E., Goldreich, P.,  
Holberg, J. P. \& Lane, A. L.} 1984b.
 The eccentric Saturnian ringlets at 
1.29 $R(s)$ and 1.45 $R(s)$.
\emph{Icarus} {\bf 60}, 1-16. 

\bibitem[Rafikov(2002)]{2002ApJ...572..566R} \textsc{Rafikov, R.~R.} 2002
 Planet Migration and Gap Formation by Tidally Induced Shocks.
\emph{ApJ} {\bf 572}, 566-579. 

\bibitem[Rosen et al.(1991)]{1991Icar...93...25R} \textsc{Rosen, P.~A., 
Tyler, G. L., Marouf, E. A. \& Lissauer, J. J.} 1991
 Resonance structures 
in Saturn's rings probed by radio occultation. II - Results and 
interpretation.
\emph{Icarus} {\bf 93}, 25-44. 

\bibitem[Shakura and Sunyaev(1973)]{1973A&A....24..337S} \textsc{Shakura, N.~I. \& 
Sunyaev, R. A.} 1973
 Black holes in binary systems. Observational appearance.
\emph{A\&A} {\bf 24}, 337-355. 

\bibitem[Smith et al.(1981)]{1981Sci...212..163S} \textsc{Smith, B.~A. \& 26 
colleagues} 1981
 Encounter with Saturn - Voyager 1 imaging science results.
\emph{Science} {\bf 212}, 163-191. 

\bibitem[Shu et al.(1985)]{shu85} \textsc{Shu, F. H., Dones, L., Lissauer, J. J., Yuan, C. \& Cuzzi, J. N.} 1985 Nonlinear spiral density waves - Viscous damping. \emph{ApJ} {\bf 299}, 542-573.
\bibitem[Strom et al.(1990)]{1990csss....6..275S} \textsc{Strom, S.~E., Edwards, S. 
\& Skrutskie, M. F.} 1990
 Evolutionary timescales for circumstellar disks 
associated with solar-type pre-main sequence stars. In \emph{ASP Conf.~Ser.~9: Cool 
Stars, Stellar Systems, and the Sun} {\bf 6}, 275-288. 

\bibitem[Tanaka and Ida(1999)]{1999Icar..139..350T} \textsc{Tanaka, H.~and Ida, S.} 
1999
 Growth of a Migrating Protoplanet.
\emph{Icarus} {\bf 139}, 350-366. 

\bibitem[Tanaka et al.(2002)]{2002ApJ...565.1257T} \textsc{Tanaka, H., Takeuchi, T. 
\& Ward, W. R.} 2002
 Three-Dimensional Interaction between a Planet and an 
Isothermal Gaseous Disk. I. Corotation and Lindblad Torques and Planet 
Migration.
\emph{ApJ} {\bf 565}, 1257-1274. 

\bibitem[Thommes et al.(1999)]{1999Natur.402..635T} \textsc{Thommes, E.~W., 
Duncan, M. J. \& Levison, H. F.} 1999
 The formation of Uranus and Neptune 
in the Jupiter-Saturn region of the Solar System.
\emph{Nature} {\bf 402}, 635-638. 

\bibitem[Thommes et al.(2002)]{2002AJ....123.2862T} \textsc{Thommes, E.~W., 
Duncan, M. J. \& Levison, H. F.} 2002a.
 The Formation of Uranus and Neptune 
among Jupiter and Saturn.
\emph{AJ} {\bf 123}, 2862-2883. 

\bibitem[]{thommes2002b} \textsc{Thommes, E.~W., Duncan, M.~J., \& Levison,
H.~F.} 2002b.
Oligarchic growth of giant planets.
Submitted to \emph{Icarus}.

\bibitem[Ward(1986)]{1986Icar...67..164W} \textsc{Ward, W.~R.} 1986
 Density waves in the solar nebula - Differential Lindblad torque.
\emph{Icarus} {\bf 67}, 164-180. 

\bibitem[Ward and Hourigan(1989)]{1989ApJ...347..490W} \textsc{Ward, W.~R. \& 
Hourigan, K.} 1989
 Orbital migration of protoplanets - The inertial limit.
\emph{ApJ} {\bf 347}, 490-495. 

\bibitem[Ward and Hahn(1995)]{1995ApJ...440L..25W} \textsc{Ward, W.~R. \& 
Hahn, J. M.} 1995
Disk tides and accretion runaway.
\emph{ApJ} {\bf 440}, L25-L28. 

\bibitem[Ward(1997)]{1997Icar..126..261W} \textsc{Ward, W.~R.} 1997
 Protoplanet Migration by Nebula Tides.
\emph{Icarus} {\bf 126}, 261-281. 



\end{thebibliography}
\end{document}